\documentclass[aps,pra,twocolumn,showpacs,preprintnumbers
longbibliography
,amsmath,amsfonts,amssymb,groupedaddress,floatfix]{revtex4-1}
\usepackage{geometry}
\usepackage[caption=false]{subfig}
\usepackage{graphicx}
\usepackage{bm}

\usepackage{amsfonts}
\usepackage{amsmath}
\usepackage{amssymb}
\usepackage{mathtools}

\usepackage[
]{hyperref}

\newcommand{\vac}{\vert {\rm vac} \rangle}

\DeclarePairedDelimiterX\braket[3]{\langle}{\rangle}%
{#1\,\delimsize\vert\,#2\,\delimsize\vert\,#3}

  \DeclarePairedDelimiterX\prodt[2]{\langle}{\rangle}%
{#1\,\delimsize\vert\,#2}

 \DeclarePairedDelimiterX\prodtt[2]{\langle \langle}{\rangle \rangle}%
{#1\,\delimsize\vert\,#2}
 
\DeclarePairedDelimiterX\brakett[3]{\langle \langle}{\rangle \rangle}%
{#1\,\delimsize\vert\,#2\,\delimsize\vert\,#3}

\DeclareSymbolFont{bbold}{U}{bbold}{m}{n}
\DeclareSymbolFontAlphabet{\mathbbold}{bbold}

\newcommand{\vect}[1]{\mathbf{#1}}

\usepackage{xkeyval}

\makeatletter
\newlength{\sfp@hseplen}\newlength{\sfp@vseplen}
\define@cmdkey{subfigpos}[sfp@]{pos}[ul]{}
\define@cmdkey{subfigpos}[sfp@]{font}[\small]{}
\define@cmdkey{subfigpos}[sfp@]{vsep}[2\baselineskip]{\setlength{\sfp@vseplen}{\sfp@vsep}}
\define@cmdkey{subfigpos}[sfp@]{hsep}[10pt]{\setlength{\sfp@hseplen}{\sfp@hsep}}
\newcommand{\subfigimg}[3][,]{%
  \setkeys{Gin,subfigpos}{pos,font,vsep,hsep,#1}
  \setbox1=\hbox{\includegraphics{#3}}
  \ifnum\pdfstrcmp{\sfp@pos}{ul}=0
    \leavevmode\rlap{\usebox1}
    \rlap{\hspace*{\sfp@hsep}\raisebox{\dimexpr\ht1-\sfp@vsep}{\sfp@font{#2}}}
    \phantom{\usebox1}
  \else\ifnum\pdfstrcmp{\sfp@pos}{ur}=0
    \leavevmode\usebox1
    \llap{\raisebox{\dimexpr\ht1-\sfp@vsep}{\sfp@font{#2}}\hspace*{\sfp@hsep}}
  \else\ifnum\pdfstrcmp{\sfp@pos}{lr}=0
    \leavevmode\usebox1
    \llap{\raisebox{\sfp@vsep}{\sfp@font{#2}}\hspace*{\sfp@hsep}}
  \else
    \leavevmode\rlap{\usebox1}
    \rlap{\hspace*{\sfp@hseplen}\raisebox{\sfp@vsep}{\sfp@font{#2}}}
    \phantom{\usebox1}
  \fi\fi\fi
}
\makeatother

\begin{document}

\title{Population dynamics in Floquet realisation of Harper-Hofstadter Hamiltonian}

\author{Thomas \surname{Bilitewski}}
\email{tb494@cam.ac.uk}
\affiliation{T.C.M. Group, Cavendish Laboratory, J.J. Thomson Avenue, Cambridge CB3 0HE, United Kingdom}
\author{Nigel R. \surname{Cooper}}
\affiliation{T.C.M. Group, Cavendish Laboratory, J.J. Thomson Avenue, Cambridge CB3 0HE, United Kingdom}

\date{\today}
\begin{abstract}
We study the recent Floquet-realisation of the Harper-Hofstadter model in a gas of cold bosonic atoms. We study in detail the scattering processes in this system in the weakly interacting regime due to the interplay of particle interactions and the explicit time dependence of the Floquet states that lead to band transitions and heating. We focus on the experimentally used parameters and explicitly model the transverse confining direction. Based on transition rates computed within the Floquet-Fermi golden rule we obtain band population dynamics which are in agreement with the dynamics observed in experiment. Finally, we discuss whether and how photon-assisted collisions that may be the source heating and band population dynamics might be suppressed in the experimental setup by appropriate design of the transverse confining potential. The suppression of such processes will become increasingly important as the experiments progress into simulating strongly interacting systems in the presence of artificial gauge fields. 
\end{abstract}


\maketitle

\makeatletter

 \newpage

\section{\label{sec:intro}Introduction}

Time-periodic driving provides a versatile tool with which to control the
behaviour of quantum systems. Such systems, described by Floquet
theory, can lead to forms of long-time dynamics that 
differ dramatically from those in the
absence of any drive. In particular, the ability to use
time-periodically modulated light fields to cause neutral atoms to
experience artificial gauge fields has sparked significant experimental
\cite{Spielman_Experiment1,Spielman_Experiment2,Jimenez,Sias_SL,Struck,Struck_SL_Ising,MagneticFields_1,MagneticFields_2,Hofstader_Bloch,mitharper,esslingerhaldane,Hofstader_Bloch_Chern}
and theoretical
\cite{Jaksch_2003,Mueller_2004,Dudarev2004,Ruseckas2005,Spielman_Theory,SOC,Kolovsky,Creffield,Flux_Lattice_Cooper_2011,Flux_Lattice_Cooper_Dalibard_2011,FQH_Cooper,Dalibard_2010,Baur_2014}
activity in studies of Floquet systems in recent years.
Whereas the long-time dynamics of a single particle in a periodically driven system can often be well understood in terms of an effective Hamiltonian \cite{Casas_2001,Rahav_2003,Bukov_2014,Goldman_2014,Goldman_2015}, the applicability of perturbative high-frequency expansions to driven interacting many-particle systems is less established \cite{Eckhardt_2015}.
Generically, under periodic drive, a closed interacting quantum system is expected to heat up to an infinite temperature state \cite{Lazarides2014,Lazarides2014a,DAlessio2014} unless many-body localisation prevents this \cite{Lazarides2014b,Abanin_2014,Ponte_2015}.
In open quantum systems, depending on the specific form of the system-bath coupling, heating to an infinite temperature state may be avoided and Gibbs-like steady states can emerge \cite{Shirai_2015,Liu_2014,Iadecola_2013}. By properly choosing the bath density of states and the coupling to the system one may obtain steady states that resemble finite temperature states of non-driven systems \cite{Deghani_2014,Deghani_2015,Iadecola_2015,Seetharam_2015}.
In either case the heating processes generically present in Floquet systems might make them unsuited to experimentally study interacting and strongly correlated phases. Thus, the need to determine the relevant timescales over which the heating processes occur becomes immediately apparent.

In a perturbative treatment of interacting Floquet systems one finds that inelastic scattering associated with the absorption of quanta $\hbar \omega$ becomes possible which can lead to heating and band transitions \cite{ScattTheoryFBS,Stability_Floquet_condensate_2}. For heating due to two-particle collisions independent of the specific lattice and driving protocol considered, it is found that the rates scale with the square of the corresponding interaction strength and scale linearly in the density of particles, and in the case of scattering into transverse dimensions can be further modified by changing the transverse density of states. Thus, appropriate design of the Hamiltonian might be used to suppress these processes, of paramount importance for experiments as deleterious heating and band transitions can limit the achievable lifetimes.
Depending on the specific setup the microscopic processes leading to such energy absorption are slightly different. In all cases they can be understood as arising from the explicit time-dependence of the Floquet-states and the non-commutativity of the interaction with the time-evolution operator of the non-interacting system \cite{ScattTheoryFBS}. If in addition to the perturbative treatment of the interactions, the time-dependence is treated approximately, e.g. in a rotating wave approximation, or the time-evolution is further expanded in small parameters, e.g. the hopping on the lattice, the heating rates can be associated with a microscopic process, e.g. a particle hopping to a lower energy site and converting that potential energy into kinetic energy, which would be possible in a time-independent system, or a genuine Floquet process in which the particles absorb energy from the driving field and convert that into energy of motion \cite{Stability_Floquet_condensate_2}. For the system we consider below it is the second process which will be of most interest, it being unique to Floquet systems. This process will naturally scale with the amplitude of the driving field, the rates typically quadratically within perturbation theory in the driving fields for small amplitudes, and as a complicated function for stronger driving. 

In this paper, we study heating processes in a Floquet system consisting of a periodically modulated superlattice potential, as used in a recent experimental realisation of the Harper-Hofstadter Hamiltonian \cite{Hofstader_Bloch_Chern}. We apply the perturbative formalism we recently developed to describe Floquet systems used to generate artificial gauge fields for cold gas systems \cite{ScattTheoryFBS}. We will treat the non-interacting time-dependent Hamiltonian exactly within Floquet theory and treat the particle-interactions perturbatively.
Focusing on the effects of two-particle scattering processes, we compute the heating processes that arise from weak two-body interactions
and corresponding intra- and inter-band transition rates, and compare with experimental observations. We further study ways in  which these processes could be suppressed by appropriate design of the Hamiltonian.

We begin by introducing the model in Sec.~\ref{sec:model} and obtain the single particle Floquet spectrum for the experimentally used parameters in Sec.~\ref{subsec:spectrum}. We proceed by introducing the Floquet scattering processes computed within the Floquet-Fermi golden rule in Sec.~\ref{subsec:scattering}. In Sec.~\ref{subsec:pop_dynamics} we compute the band population dynamics based on the Floquet scattering rates for the experimentally used parameters. Finding reasonable agreement with the experimentally observed dynamics we then discuss in Sec.~\ref{subsec:stab_scatt_rates} whether these processes can be suppressed by appropriate design of the transverse confining potential. We show that a strong suppression can be achieved under sufficiently strong confinement, suggesting a possible way to enable experiments to access strongly correlated quantum phases without the deleterious heating and repopulation dynamics demonstrated to be present in this Floquet system at weak confinement.

\section{\label{sec:model}Model}

We consider bosons described by a field-operator $\Psi(\vect{x})$ loaded into a two-dimensional (2D) optical lattice with both time-dependent and static parts with additional transverse confinement. The resulting Hamiltonian is given by $H=H_0(t) + H_{\text{int}}$,
\begin{align}
  \begin{split}
 H_0(t)&= \int d\vect{x}  \,\frac{- \hbar^{2}}{2m} \Psi^{\dagger}(\vect{x}) \Delta \Psi(\vect{x})  \\
  & +\int d\vect{x}  \, \Psi^{\dagger}(\vect{x})\left[ V(x,y,t) +V^{z}(z) \right] \Psi(\vect{x}) ,   
  \end{split} \label{eq:lattice_H0}\\
 H_{\text{int}} &=\frac{g}{2} \int{d\vect{x} \, \Psi^{\dagger}(\vect{x}) \Psi^{\dagger}(\vect{x}) \Psi(\vect{x}) \Psi(\vect{x})} \label{eq:Lattice_Hint},
\end{align}
where the first line of equation~(\ref{eq:lattice_H0}) describes the kinetic energy of the atoms moving in the optical lattice and the second line gives the time-dependent in-plane optical potential $V(x,y,t)$ and the static transverse confining potential $V^{z}(z)$ experienced by the atoms.
The last part, equation~(\ref{eq:Lattice_Hint}), gives the contact interaction between atoms of strength $g$. For the weak contact interactions considered here, $g = 4\pi \hbar^2 a_{\rm s}/M$ describes collisions with  $s$-wave scattering length $a_{\rm s}$.

We consider the in-plane optical lattice $V(x,y,t)$ to be a dynamical superlattice\cite{Baur_2014}, which for the experimental setup of Ref.~\onlinecite{Hofstader_Bloch_Chern} is given by
\begin{equation}
  \begin{split}
 V(x,y,t)&=V_{\rm st}(x,y) + V_{\rm osc}(x,y,t) \\
         &=V_x \sin^2(k_r x) + V_{xl} \sin^2(k_r x/2 ) + V_y \sin^2(k_r y) \\
         &\quad +\kappa \left[\sin(\pi/4 + k_r x/2) \cos(\phi_0 + \omega t - k_r y/2) \right.\\
         &\quad \quad \left. -\cos(\pi/4 + k_r x/2)  \sin(\phi_0 - \omega t- k_r y/2)  \right]
  \end{split}\label{eq:lattice_potential}
\end{equation}
where the second line describes the static super lattice $V_{\rm st}(x,y)$, consisting of the short lattice in $x (y)$ direction of strength $V_x (V_y)$ and the long lattice with strength $V_{xl}$ that creates the staggering along the $x$-direction. The third and fourth line describe the time-dependent part $V_{\rm osc}(x,y,t) $ of strength $\kappa$ oscillating with frequency $\omega$ which is fixed to be on resonance with the energy staggering along the $x$-direction. The time-dependent part is chosen such that the second and third line respectively restore tunnelling along each other bond in the staggered $x$-direction, and such that a homogeneous flux of $\pi/2$ is realised in the tight-binding description. The phase $\phi_0$ is not controlled in the experiment and will be fixed in the following to be $\phi_0=\pi/4$; we have checked that the results are not changed significantly for a different choice.
The transverse confinement potential is taken to be either an optical lattice $V^{z}_{\rm lat}=V_z \cos^2(k^{z}_r z)$ or a harmonic trap $V^{z}_{\rm osc} = \frac{1}{2} m\omega_{\rm osc} z^2$ which is used in the experiment.

\subsection{\label{subsec:spectrum}Single-particle states and dispersion}
\begin{figure}
 \includegraphics[width=0.48\textwidth]{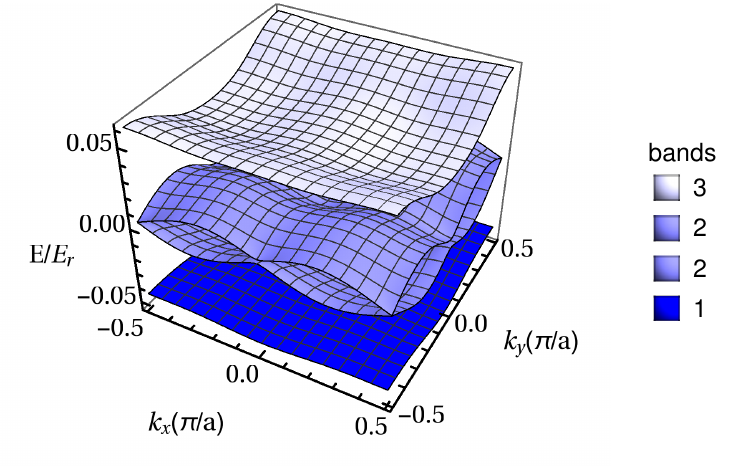}
    \caption{(Color online) Resulting band structure $E_n(k_x,k_y)$ in units of the recoil-energy $E_r=\hbar^2 k_r^2/(2M)$ as a function of momentum $k_x$ and $k_y$ in the reduced BZ of the non-interacting model, Eq.~(\ref{eq:lattice_H0}), for the parameters $V_x=6 E_r$, $V_{xl} = 0.8 E_r$, $V_y=10 E_r$ and $\kappa = 0.58 \hbar \omega$ and $\hbar \omega = 0.72 E_r$. The dispersion splits into 4 bands of which the middle two bands are touching at the borders of the reduced BZ and combined into a single superband.}
    \label{fig:dispersion}
\end{figure}

We proceed to obtain the single-particle Floquet spectrum of the non-interacting time-dependent Hamiltonian $H_0(t)$. Due to the separability of the non-interacting Hamiltonian we may focus on the particle spectrum of the in-plane ($x$-$y$) motion, i.e. wavefunctions take the form $\Psi(x,y,z,t)=\Psi^{\rm 2D}(x,y,t) \Psi^{z}(z)$ and energies are given by $E_{n,k,m^{z}}=E^{\rm 2D}_{n,k}+E^{z}_{m^z}$. As the Hamiltonian is invariant under discrete temporal and spatial translations the solutions take the form of Floquet-Bloch waves. 

We note that the static potential has the following symmetries, $V_{\rm st}(x+2 a,y)=V_{\rm st}(x,y+a)=V_{\rm st}(x,y)$ with the lattice spacing $a=\pi/k_r$. We may choose a unit cell of $2\times 2$ sites and each Bloch-band will be split into 4. However, the time-dependent part has a lower symmetry, $V_{\rm osc}(x+4a,y,t)=V_{\rm osc}(x,y+4a,t)=V_{\rm osc}(x,y,t)$, and naively, this would lead to a real space unit cell of $4 \times 4$ sites and to a splitting into 16 bands. However, the unit cell can be reduced by rewriting $V_{\rm osc}$ as
\begin{equation}
 V_{\rm osc}(x,y,t) = e^{i \omega t} F(x,y) + e^{-i \omega t} F^{*}(x,y)
\end{equation}
with quasi-periodic $F(\vect{r} +\vect{R_j})= e^{i \vect{G} \cdot \vect{R_j}}F(\vect{r})$ with $\vect{G}=(\pi/(2a),\pi/(2a))$. This allows us to perform a unitary gauge-transformation in Floquet-space as done in \cite{Baur_2014} and obtain a Hamiltonian invariant under translations by 2 lattice sites in both $x$- and $y$-directions. Consequently, we may keep the unit cell consisting of $2\times2$ sites with each band split into 4 and exactly expand the time-dependent problem in the Bloch-states of the time-independent problem as they now share the same periodicity.

We project the full time-dependent Schr\"odinger equation on the set of the 4 lowest Bloch-bands of $V_{\rm st}$ which are resonantly coupled by the time-dependent optical potential $ V_{\rm osc}$. Moreover, in the expansion of the Floquet-states, $\Psi_{\epsilon}(t)=e^{i \epsilon t} \Phi_{\epsilon}(t) = e^{i \epsilon t}  \sum_m \phi_m e^{i m \omega t}$, we keep only a finite number of frequency components $-M \le m \le M$ with sufficiently high $M$ to ensure convergence of both the spectrum and wavefunctions.

A plot of the resulting band structure is shown in Fig.~(\ref{fig:dispersion}) which shows the 4 Harper-Hofstadter bands in the reduced Brillouin zone corresponding to the $2\times2$ real-space unit-cell. Following the convention in \cite{Hofstader_Bloch} we refer to the middle two bands as a single band as they are not separated by an energy gap.

\subsection{\label{subsec:scattering}Floquet-Scattering}
\begin{figure}
 \includegraphics[width=0.49\textwidth]{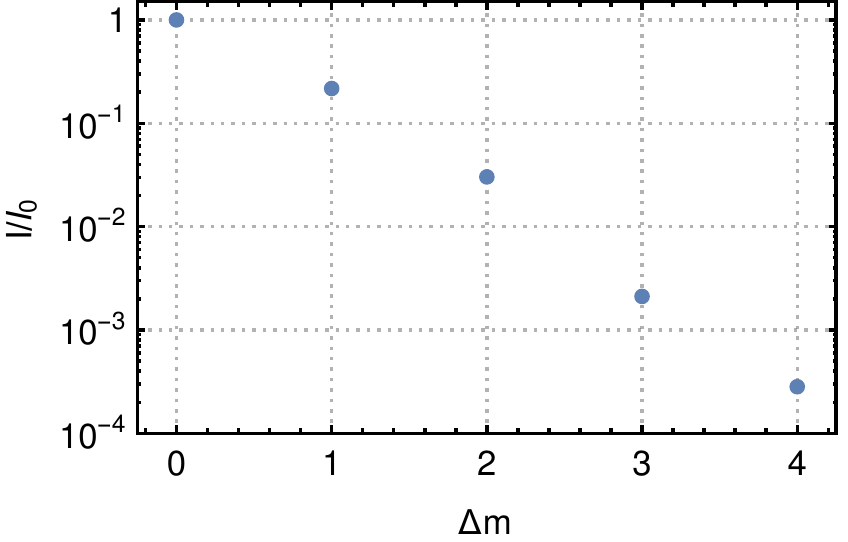}
    \caption{(Color online) Overlap matrix element $I_{\Delta m}=|\brakett{\Phi_{\rm f}^{\Delta m}(x,y)}{H_{\text{int}}}{\Phi_{\rm i}^{0}(x,y)}|^2$ of the wavefunctions of the 2 dimensional model for a transition $(1,1)\rightarrow(3,3)$ summed over the final state momenta and averaged over the initial state momenta as a function of the photon transfer $\Delta m$ normalised to the element at $\Delta m =0$.}
    \label{fig:trans_rates_m}
\end{figure}

We treat the effects of the collisions of the atoms given by the interaction Hamiltonian~(\ref{eq:Lattice_Hint}) in the framework of scattering theory for Floquet states, \cite{ScattTheoryFBS} and references therein. We employ  the Floquet-Fermi golden rule (FFGR) to compute transition rates from an initial state $\Psi_{\rm i}$ to final state $\Psi_{\rm f}$ given by
\begin{equation}
 \gamma_{\rm i\rightarrow \rm f} =\sum_{\Delta m}  \frac{2 \pi}{\hbar} \delta(\epsilon^0_{\rm i}-\epsilon^0_{\rm f}-\Delta m\hbar \omega) |\brakett{\Phi_{\rm f}^{\Delta m}}{H_{\text{int}}}{\Phi_{\rm i}^{0}}|^2 
\label{eq:FFGR}
\end{equation}
where $\Phi^0_{{\rm (i/f)}}$ is the time-periodic Floquet mode associated with $\Psi_{{\rm (i/f)}}$, $\Phi^m(t)=e^{i m \hbar \omega t} \Phi^{0}(t)$ and the double brackets denote the extended scalar product given by $\prodtt{\Phi_{\rm f}}{\Phi_{\rm i}}=1/T \int_0^{T} \prodt{\Phi_{\rm f}(t)}{\Phi_{\rm i}(t)} $. The sum over $\Delta m$ allows the absorption or emission of $\Delta m$ energy quanta $\hbar \omega$ during the scattering process. 

Using the FFGR will preclude the discussion of strongly correlated many-body phases, but is appropriate for the regime of weakly-interacting particles considered here and sufficient to explore the two-particle physics that are a potential source of heating and band repopulation. 

We apply FFGR Eq.~(\ref{eq:FFGR}) to an initial state containing particles in bands $n_1$ and $n_2$ with in-plane crystal-momentum $k_1$ and $k_2$ of the 2D bandstructure in states $m_1^{z}$ and $m_2^{z}$ of the transverse potential, i.e.
\begin{equation}
 \Psi_{\rm i}=\Psi^{a \dagger}_{n_1,k_1,m_1^{z}} \Psi^{a \dagger}_{n_2,k_2,m_2^{z}}\vac 
\end{equation}
scattering into a final state containing two particles in bands $n_3$ and $n_4$ with crystal-momentum $k_3$ and $k_4$ in states $m_3^{z}$ and $m_4^{z}$
\begin{equation}
  \Psi_{\rm f}=\Psi^{a \dagger}_{n_3,k_3,m_3^{z}} \Psi^{a \dagger}_{n_4,k_4,m_4^{z}}\vac .
\end{equation}

In the calculations of the population dynamics in Sec.~\ref{subsec:pop_dynamics} we only keep transitions with emission and absorption of up to a single energy quantum $\hbar \omega$, i.e. $\Delta m=0,\pm 1$ in the FFGR Eq.~(\ref{eq:FFGR}), as transitions with higher energy transfers are successively suppressed as shown in Fig.~(\ref{fig:trans_rates_m}). Note that we still have to keep a high number $M$ of Floquet modes in the expansion of the Floquet states used to compute the rates. The exponential decay of these transition-matrix elements also justifies our restriction to the lowest 4 Harper-Hofstadter bands. Whereas transitions into the higher bands are allowed within FFGR, for the experimental parameters they correspond to a transition with $\Delta m \approx 11$ and are thus negligible on the time-scales we are interested in. This argument applies rather generally and does not rely on the precise form of the wavefunctions, but only on the fact that the higher bands are gapped in the static Hamiltonian and that the Floquet states are exponentially localised in the frequency domain, thus, leading to the observed exponential suppression of the matrix elements.

For the general discussion of the scattering rates in Sec.~\ref{subsec:stab_scatt_rates} we assume an initial state in the unique lowest energy state of the transverse direction, i.e. $m_1^{z}=m_2^{z}=0$ for harmonic confinement, and  define a total dimensionless rate for transitions $\Gamma_{(n_1,n_2)\rightarrow (n_3,n_4)}$ taking the initial state to be $({n_1,k_1,0; n_2,k_2,0})$, summing over the final state momenta $k_3$ and $k_4$ and transverse state quantum numbers $m_3^{z}$ and $m_4^{z}$ and averaging the initial state momenta $k_1$ and $k_2$ over the reduced Brillouin zone (BZ)
\begin{equation}\label{eq:band_rate}
\begin{split}
  \Gamma_{(n_1,n_2)\rightarrow (n_3,n_4)} \gamma^{0} &= \sum_{m_3^{z},m_4^{z}}\left(\frac{2a}{\pi}\right)^4\left(\frac{L_x L_y}{(2 \pi)^2}\right)^2 \\
                                                     &\quad \iint_{BZ} d\vect{k}_1 d\vect{k}_2 d\vect{k}_3 d\vect{k}_4  \, \gamma_{\rm i\rightarrow \rm f}
  \end{split}
\end{equation}
where we factored out the dimensionful prefactor $\gamma_0$. The prefactor $\gamma_0$ takes the form 
$\gamma_0=\frac{g^{2}}{\hbar E_r} k_r^{2} \frac{1}{L_x L_y} \frac{E_r}{\hbar \omega_{\text{osc}}} \frac{m \omega_{\text{osc}}}{\hbar }  $ for harmonic confinement and  
$\gamma_0=\frac{g^{2}}{\hbar E_r} k_r^{2} \frac{1}{L_x L_y L_z}  \frac{E_r}{E_r^{z}} k_r^{z} $ for confinement by an optical lattice.

\subsection{\label{subsec:pop_dynamics}Population dynamics}
We compute the resulting band population dynamics based on the Floquet transition rates, Eqn.~(\ref{eq:FFGR}), for the experimental parameters, i.e. $V_x=6 E_r$, $V_{xl} = 0.8 E_r$, $V_y=10 E_r$, $\kappa = 0.58 \hbar \omega$, $\hbar \omega = 0.72 E_r$ and $E_r=\frac{\hbar^2 k_r^{2}}{2m} $ with $k_r=\pi/a$ and $a=0.5\times 767$ nm, and for the experimentally used harmonic confinement  $V^{z}_{\rm osc} = \frac{1}{2} m\omega_{\rm osc} z^2$ with $\omega_{\text{osc}}=  12 $Hz. 
We use the value of the scattering length for $^{87}$Rb and the experimental density for which $N d^2/(L_x L_y) \approx 20$. We assume the particles to be homogeneously spread over the BZ and average all rates over the initial state momenta as discussed above. This allows us to formulate the problem only in terms of the band-populations of the 2D Harper-Hofstadter model and the transverse direction.
\begin{figure}
 \includegraphics[width=0.49\textwidth]{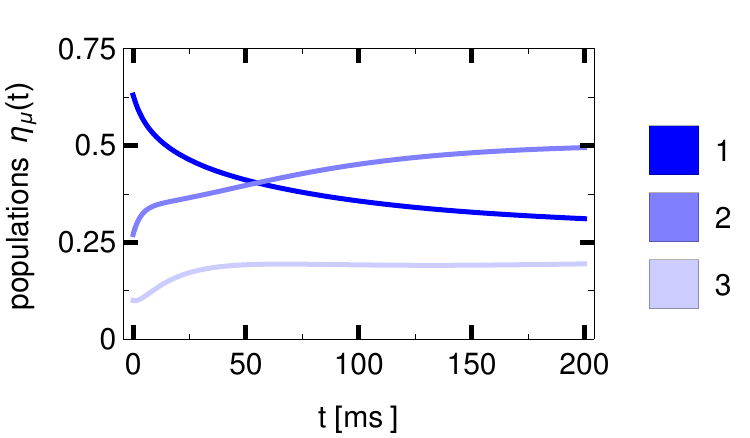}
    \caption{(Color online) Dynamics of the fractional band populations $n_{\mu}=\sum_{m^{z}} N_{\mu,m^{z}}/N_{\text{total}}$ in the Floquet realization of the Harper-Hofstadter model for the experimental parameters with a transverse harmonic confinement summed over the $m^{z}$ oscillator quantum number. Colours correspond to the bands shown in Fig.~\ref{fig:dispersion}.}
    \label{fig:band_dynamics}
\end{figure}
The resulting dynamics for the fractional band occupations $n_{\mu}=\sum_{m^{z}} N_{\mu,m^{z}}/N_{\text{total}}$ summed over the states of the transverse confining harmonic potential are shown in Fig~\ref{fig:band_dynamics} for initial fractional occupations of $n_1=0.6$, $n_2=0.3$ and $n_3=0.1$, all assumed to be initially in the $m^{z}=0$ state. The behaviour compares favourably with the experimental results, see Fig. 3 in \cite{Hofstader_Bloch_Chern}.

We observe two time scales for the population dynamics. On a fast time-scale elastic ($\Delta m=0$) collisions of the type $(1,3)\rightarrow (2,2)$ and the reverse process redistribute particles between the bands while conserving the in-plane motion energy, i.e. the transverse motion quantum numbers $m_i^{z}=m_f^{z}$ are conserved. As expected from Fig.~\ref{fig:trans_rates_m} these occur considerably faster than the Floquet $\Delta m \ne 0$ transitions. This initally leads to an increase of the population in band 2 and the corresponding decrease of populations in band 1 and 3, while keeping all particles in the $m^{z}=0$ states of the transverse potential. For the initial conditions, these processes achieve a quasi-equilibrium on a very short time-scale stopping further redistribution. For longer times Floquet transitions ($\Delta m=\pm 1$), changing both the Hofstadter bands and allowing transfer into higher energy states of the tranverse direction $m_i^{z} \ne m_f^{z}$, become relevant. The interplay of the elastic and the inelastic collisions is seen to reverse the initial decrease in population of band 3. As the higher states of the transverse potential are occupied the reverse scattering processes become important and slow down the population dynamics after a time scale of about $10$ ms. The further dynamics is seen to be slower explained by the fact that the higher states of the transverse potential are less strongly interacting. Finally, the system approaches a state in which all Harper-Hofstadter bands are equally populated over a timescale of $200$ ms. 

\begin{figure}
 \includegraphics[width=0.49\textwidth]{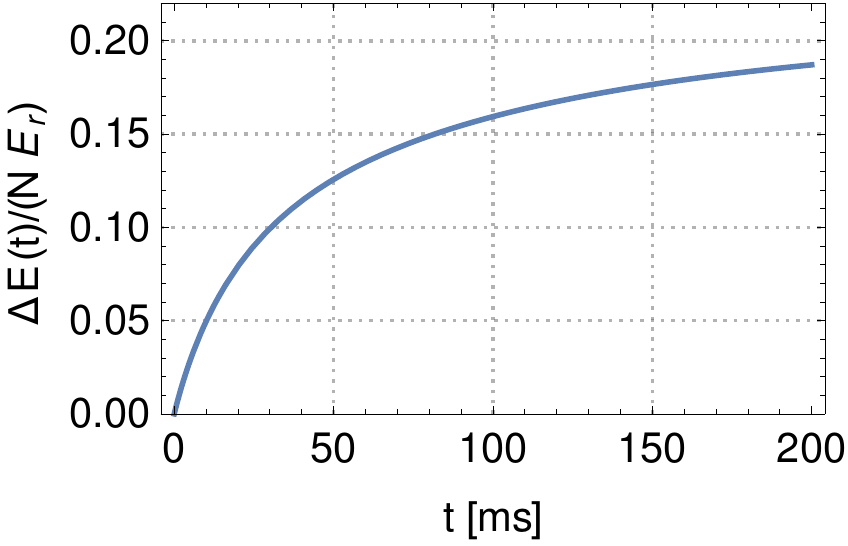}
    \caption{(Color online) Total absorbed energy $\Delta E(t)= E(t)-E(t=0)$ per particle in units of $E_r$ in the Floquet realization of the Harper-Hofstadter model for the experimental parameters with a transverse harmonic confinement. }
    \label{fig:transverse_energy}
\end{figure}

We remark that even though the Harper-Hofstadter band populations approach a steady state, the system slowly continues to heat up in the transverse direction. The total absorbed energy $\Delta E(t)= E(t)-E(t=0)$ per particle in units of $E_r$ is shown in Fig.~\ref{fig:transverse_energy}. After an initial period of fast energy absorption the rate flattens out. This is due to the fact that the interactions become weaker as the particles are transferred to higher $m^{z}$ states. In the beginning all particles occupy $m^{z}=0$ states and are strongly interacting.
The $\Delta m=\pm 1$ photon transitions then establish an equilibrium between states at $m^{z}=0$ and $m^{z}=m^{z}_{\omega}=\hbar \omega/E_{osc}$. Consequently, particles in the $m^{z}_{\omega}$ states would be scattered to states in the  $m^{z}=2 m^{z}_{\omega}$ state. The timescale for this to happen is controlled by the square of the ratio of the matrix elements $\braket{\Psi^{z}_{m^{z}_{\omega}/2} \Psi^{z}_{m^{z}_{\omega}/2}}{H_{int}}{\Psi^{z}_{m^{z}_{\omega}} \Psi^{z}_{m^{z}_{\omega}} }$ and $\braket{\Psi^{z}_0 \Psi^{z}_0 }{H_{int}}{\Psi^{z}_{m^{z}_{\omega}/2} \Psi^{z}_{m^{z}_{\omega}/2} }$ which is approximately $0.06$, yielding transition rates 3 orders of magnitude lower. 
We observe that the total absorbed energy per particle remains small compared to the bandgaps of the undriven system, thus, being consistent with only considering the 4 lowest bands for these time scales.

We conclude that the strong inter band population dynamics and the associated heating stemming from $\Delta m=\pm 1$ transitions would make the realization of strongly interacting phases exceedingly difficult unless proper care is taken to suppress the corresponding processes.

\subsection{\label{subsec:stab_scatt_rates}Stability and scattering rates}
Based on the conservation of quasi-energy in the FFGR Eq.~(\ref{eq:FFGR}) and the specific experimental parameters, one can readily envisage ways in which the transverse density of states can be adapted in order to make the system stable to
the two-particle scattering processes with $\Delta m =\pm 1$ which cause band transfer and lead to heating.

\begin{figure} 
\centering 
\subfloat{\subfigimg[pos=ul,hsep=0.0\textwidth,vsep=0.0\textwidth,width=0.45\textwidth]{\bf (a)}{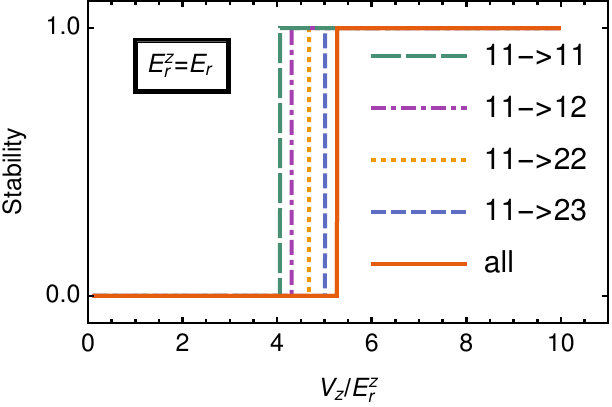}}
\hfill
\subfloat{\subfigimg[pos=ul,hsep=0.0\textwidth,vsep=0.0\textwidth,width=0.45\textwidth]{\bf (b)}{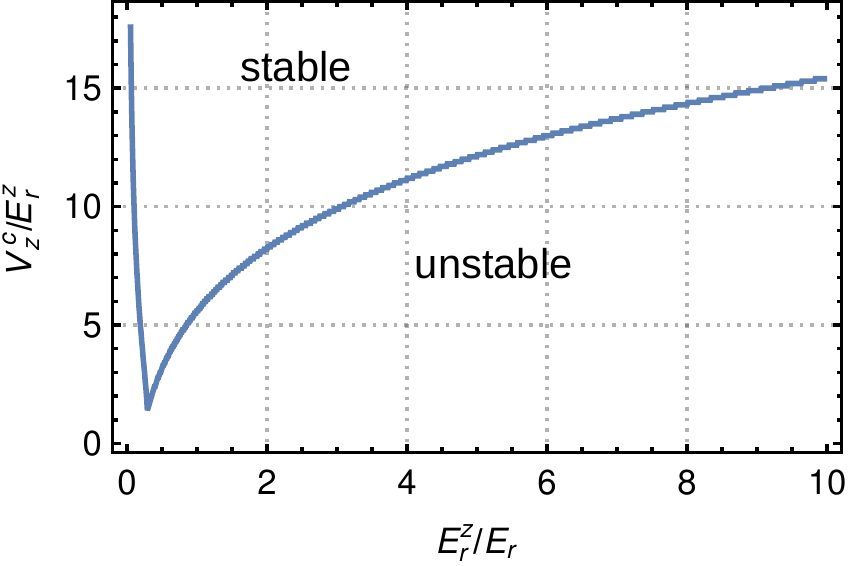}} 
\caption{(Color online) In (a) stability of the groundstate band of the Harper-Hofstadter model to different inelastic transitions as indicated in the figure in the case of transverse confinement by an optical lattice $V^{z}_{\rm lat}=V_z \cos^2(k^{z}_r z)$ of strength $V_z/E_r^{z}$ with $E_r^{z}=E_r$. A jump from 0 to 1 indicates the change of allowed to forbidden for the respective transition. In (b) the critical confinement strength $V_z^{c}/E_r^{z}$ required to suppress the $\Delta m=\pm 1$ transitions as a function of $E_r^{z}/E_r$.}
 \label{fig:trans_lat_stability}
\end{figure}

We consider the cases in which the transverse potential in the $z$-direction is either an optical lattice $V^{z}_{\rm lat}=V_z \cos^2(k^{z}_r z)$ or a harmonic trap $V^{z}_{\rm osc} = \frac{1}{2} m\omega_{\rm osc} z^2$. 
For both cases we investigate whether particles in the ground state band\footnote{while generically there is no preferred ordering of the bands in the Floquet BZ and therefore no ground state we use this notion based on the character of the bands in terms of bands of the undriven Hamiltonian} of the Harper-Hofstadter model can possibly undergo transitions with $\Delta m =\pm 1$ depending on the strength of the transverse potential. The resulting stability diagram in the case of the optical lattice is shown in Fig.~(\ref{fig:trans_lat_stability}) and in the case of the harmonic trap in Fig.~(\ref{fig:trans_osc}). This stability is based purely on kinematic constraints of the scattering process and is therefore valid for two-particle scattering beyond the applicability of the FFGR. We conclude that by sufficiently strong transverse potential the inelastic 2 particle single-photon-transfer scattering processes can be completely suppressed.

%

We begin by discussing the case in which the transverse potential is an optical lattice. For vanishing depth, $V_z=0$, we recover the situation of free particles where the system can always absorb energy and is therefore unstable. As we increase the lattice depth $V_z$ the transition with the smallest energy transfer, the intraband scattering transition $(1,1)\rightarrow (1,1)$, is suppressed first, and those with higher energy transfer are consecutively suppressed until finally the last transition, $(1,1)\rightarrow (3,3)$, becomes energetically forbidden and the system is stable to single photon ($\Delta m=\pm 1$) transitions. This happens when the density of states in the transverse direction vanishes for the required energy transfer, i.e. for these parameters when the bandwidth of the lowest band in the transverse direction becomes smaller than $\hbar \omega -2 [\max(E_3)-\min(E_1)]$. 
Additionally, we may change the ratio $E_r^{z}/E_r$ which determines the critical value of the confinement strength as shown in the right panel of Fig.~(\ref{fig:trans_lat_stability}) which is minimised at $V_z=1.5 E_r^{z}$ for $E_r^{z}/E_r= 0.36$.

\begin{figure} 
\centering 
\subfloat{\subfigimg[pos=ul,hsep=0.0\textwidth,vsep=0.0\textwidth,width=0.45\textwidth]{\bf (a)}{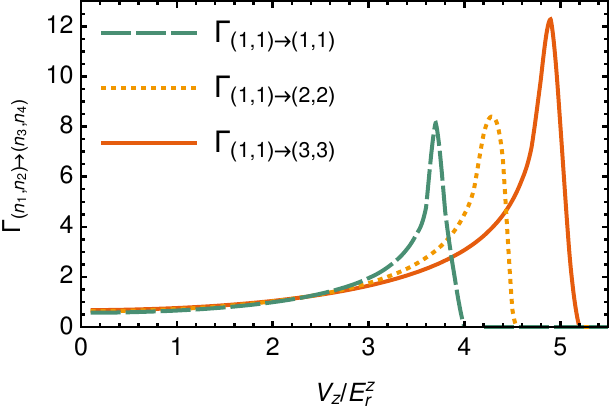}}
\hfill
\subfloat{\subfigimg[pos=ul,hsep=0.0\textwidth,vsep=0.0\textwidth,width=0.45\textwidth]{\bf (b)}{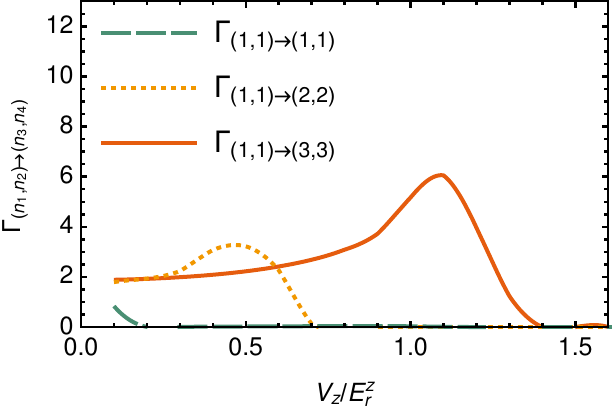}} 
    \caption{(Color online) Scattering rates $\Gamma_{(n_1,n_2)\rightarrow (n_3,n_4)}$ for the band indicated in the figure in the case of transverse confinement by an optical lattice $V^{z}_{\rm lat}=V_z \cos^2(k^{z}_r z)$ for $E_r^{z}=E_r$ in panel (a) and $E_r^{z}= 0.36 E_r$ in (b) corresponding to the minimum of $V_z^{c}$ observed in Fig.~\ref{fig:trans_lat_stability}.}
    \label{fig:trans_lat2}
\end{figure}
%

The corresponding transition rates for $E_r^{z}=E_r$ are shown in Fig~(\ref{fig:trans_lat2}). These start from the finite non-zero value at $V_z=0$ expected for free particles, and increase with increasing $V_z$ to reach a maximum. This maximum occurs at the point at which there is a closing of the channel for scattering into higher transverse modes, and arises from the characteristic singular density of states of the transverse lattice at the edges of the BZ. Thus, we see that the transverse optical lattice at first increases the scattering rates until it finally completely suppresses two-particle single-photon scattering for sufficiently deep lattices.
In addition to the lattice depth $V_z/E_r^{z}$ which determines the form of the density of states in the transverse direction, changing the ratio $E_r^{z}/E_r$ which effectively rescales the energy axis of the transverse direction compared to the 2D Harper-Hofstadter model can be used to control the scattering rates. This effect can be understood by the fact that at fixed $\hbar \omega$ reducing $E_r^{z}$ requires the transitions to couple to a final state containing particles in higher Bloch bands of the transverse lattice which have smaller overlap with the ground state Bloch band which is assumed to be the initial state. This leads to the decrease of the peak scattering rate as seen in panels (a) and (b) in Fig.~\ref{fig:trans_lat2}.

For the harmonic trap the behaviour is slightly more complicated as seen in Fig~(\ref{fig:trans_osc}). Again for no transverse potential, $E_{\rm osc} =\hbar \omega_{\rm osc} =0$, we recover the behaviour of the free system. As we increase the confinement energy $E_{\rm osc}$ transitions become possible and forbidden whenever the required transition energy matches an integer number of the oscillator energy. For weak confinement, this typically includes a range of possible final oscillator states for a given band-transition which shrinks with increasing confinement. The system is stable to single-photon ($\Delta m=\pm 1$) transitions when the energy balance equation $2n \hbar \omega_{\rm osc} =\hbar \omega -E_{\rm trans}$ has no solution with integer $n$ for all band transitions. Here we obtain two stability regions, one in the intermediate regime around $\omega_{\rm osc} \approx \omega/3$ and one for strong confinement $\omega_{\rm osc} \gtrapprox  \omega/2$.

\begin{figure} 
\centering 
\subfloat{\subfigimg[pos=ul,hsep=0.0\textwidth,vsep=0.0\textwidth,width=0.45\textwidth]{\bf (a)}{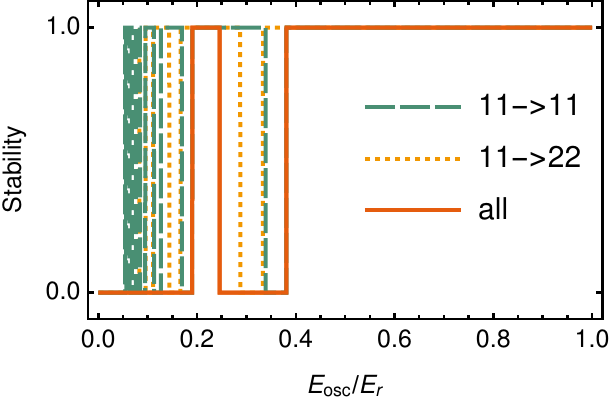}}
\hfill
\subfloat{\subfigimg[pos=ul,hsep=0.0\textwidth,vsep=0.0\textwidth,width=0.45\textwidth]{\bf (b)}{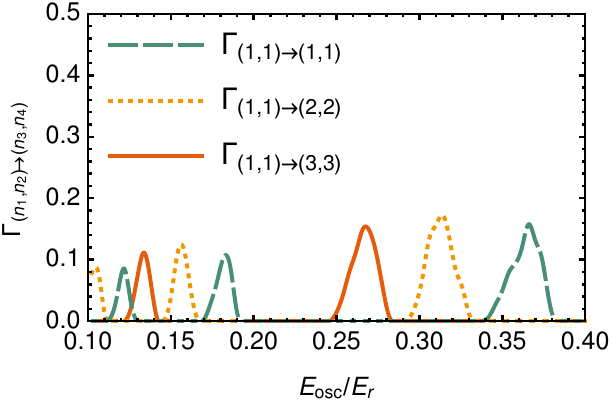}} 
    \caption{(Color online) Stability of the groundstate band of the Harper-Hofstadter model to different inelastic transitions as indicated in the figure in the case of transverse confinement by a harmonic potential $V^{z}_{\rm osc} = \frac{1}{2} m\omega_{\rm osc} z^2$ in panel (a) and corresponding scattering rates in panel (b) both as a function of the oscillator energy $E_{\rm osc} = \hbar \omega_{\rm osc}$ in terms of the recoil energy $E_r$.}
    \label{fig:trans_osc}
\end{figure}

In both cases, by sufficiently strong modifications of the transverse density of states,
 the single photon ($\Delta m=\pm 1$) transitions can be completely suppressed. However, in that regime the next to leading order processes of transitions with $\Delta m=\pm 2$ may still be present. Due to the localisation of the Floquet modes observed in the matrix elements seen in Fig.~\ref{fig:trans_rates_m} these rates are at least an order of magnitude smaller than the $\Delta m=\pm 1$ rates. In addition they are suppressed by the interaction matrix element of the wavefunctions of the transverse direction for large energy transfer, e.g. in the case of harmonic confinement we have $\braket{\Psi^{z}_{0} \Psi^{z}_{0}}{H_{int}}{\Psi^{z}_{m^{z}_{\omega}} \Psi^{z}_{m^{z}_{\omega}} }/\braket{\Psi^{z}_0 \Psi^{z}_0 }{H_{int}}{\Psi^{z}_{0} \Psi^{z}_{0} }\approx 0.06 $, yielding 3 orders of magnitude compared to the elastic ($\Delta m=0$) and a factor of 4 compared to the $\Delta m=\pm 1$ rates. In particular, the elastic scattering rates will then be at least two orders of magnitude higher than the first non-vanishing Floquet rates.

\section{\label{sec:conclusions}Conclusions}
We have studied the two-particle scattering processes occurring in the Floquet-realisation of the Harper-Hofstadter Hamiltonian focusing on the experimental setup used in \cite{Hofstader_Bloch_Chern} involving a dynamically modulated superlattice potential. Using the Floquet-Fermi golden rule we compute scattering rates due to particle interactions. Based on these Floquet transition rates we obtain the resulting band population dynamics and compare to the experimental results of \cite{Hofstader_Bloch_Chern}. The agreement between the experimental result and the predictions of this rate model suggests that two-particle processes and the Floquet scattering rates might play an important role in establishing the band dynamics and the heating rates in this system. 
Of
particular relevance to the stability of a Bose-Einstein condensate
(BEC) in one of the Harper-Hofstadter bands, we observe that the
scattering rates do not depend strongly on the final state
momenta. Consequently, an initial BEC quickly spreads over the BZ and into other subbands.

We remark that in a different experimental realisation of the Harper-Hofstadter model \cite{MIT_HARPER_2015} heating processes have been investigated very carefully and two-particle scattering was found not to be the limiting factor for the lifetimes in that setup.

More generally, our study provides insight into the heating dynamics of
a closed quantum system with an unbounded dispersion subject to
periodic driving. We find a timescale over which the system approaches
an infinite temperature state for the bounded degrees of freedom of
the in-plane motion and a generically different time-scale over which
the system then continues to heat up in the transverse direction.

Having established these processes in the Floquet system with the geometry of Ref.\onlinecite{Hofstader_Bloch_Chern}, we discussed how these rates can be influenced by additional transverse confinement. We studied transverse confinement by both an optical lattice and a harmonic potential and concluded that by choosing a sufficiently strong confinement inelastic single photon transitions with the absorption or emission of $\hbar \omega$ can be suppressed completely. Moreover, in the case of confinement by an optical lattice, the scattering rates can be further controlled by choosing an optimal values for the ratio $E_r^{z}/E_r$ in addition to the lattice depth $V^{z}$. 
By suppressing the single-photon $\Delta m=\pm 1$ rates one can achieve a regime in which the next order Floquet processes $\Delta m=\pm 2$ are at least two orders of magnitude smaller than the elastic $\Delta m=0$ rates which are responsible for establishing the strongly correlated behaviour of the quantum system.
This possibility to strongly suppress the two-particle inelastic scattering rates provides one possible route towards the design of future experiments aiming to access strongly interacting regimes without deleterious scattering and heating.

\begin{acknowledgments}
We thank the anonymous referee for constructive and helpful comments. We are grateful to Monika Aidelsburger and Immanuel Bloch of discussions. This work was supported by EPSRC Grant No EP/K030094/1.
\end{acknowledgments}

\bibliography{bib.bib}{}

\end{document}